\documentclass{article}

\usepackage{arxiv}

\usepackage[utf8]{inputenc} 
\usepackage[T1]{fontenc}    
\usepackage{hyperref}       
\usepackage{url}            
\usepackage{booktabs}       
\usepackage{amsfonts}       
\usepackage{nicefrac}       
\usepackage{microtype}      
\usepackage{lipsum}
\usepackage{graphicx}

\title{Keplers's Equation and Angular Momentum: Historical Perspective, Critical Analysis and Implications for Development of the  Orbital Mechanics/Dynamics, Mathematics and Physics}

\author{
 Slobodan Nedi\'{c} \\
  University of Novi Sad, Faculty of Technical Sciences\\
  Trg Dositeja Obradovi\'{c}a 6, 21000 Novi Sad, SERBIA \\
  \texttt{nedics@uns.ac.rs ; nedic.slbdn@gmail.com} \\
}

\begin{document}

\maketitle
\begin{abstract}
	\noindent After some more than four centuries from the formulation and publication (in \textit{Astronomia Nova}) of the  Kepler's Equation, which relates the eccentric (and, intermediately, the true) anomaly of the planetary trajectories to the uniformly flowing time, in accordance with his Second (``Area'') law, the subsequently -- in course of development of Orbital Mechanics -- to the 2${}^{nd}$ law related and formally derived non-existent (zero-valued)  transverse acceleration is questioned. Certain implications to Elliptic Integration, Symplectic Integration, Symplectic Geometry/Topology, as well as the connection between physical and mathematical continua in the context of the multi-level, scale-invariant mechanics/dynamics (with the augmented central and torquing forces) are also briefly hinted to.
\end{abstract}


\section{Introduction}
Along the total energy consisted of the sum of kinetic and potential energies, it is the angular momentum ($L$ = $r^{2} \dot{\varphi }$) of an orbital body that has been taken as time-invariant during the whole orbital period (\textit{T}), thus forming the so-called (the two) Prime Integrals for the process of solving the related (non-oscillatory) non-linear differential equations of motion. While the Energy Integral can quite straightforwardly be shown to have been untenable a postulate, unless adopting dependability on the zero-valuedness of the transverse acceleration \footnote{\ Quite\ rare\ textbooks'\ authors\ attempt\ to\ prove\ its\ validity\ either\ through\ presumed\ co-linearity\ of\ the\ position\ and\ velocity\ vectors,\ essentially\ unfulfilled\ either\ in\ initial\ conditions\ or\ in\ the\ produced\ solutions.} (either by direct analytical and/or numerical evaluation (App. A in article's extended version, with link in Footnote 4), and/or by application of the full time derivative of the related expression), for the angular momentum -- and it's Kepler's Second (Area) Law counterpart -- that turns out to be quite subtle and very intricate an issue. However, due to the very presence of non-zero tangential acceleration in the observed trajectories of orbital bodies as well as in the produced solutions, the validity of the related First Integral `status' of this quantity should have long been questioned because the constant angular momentum\footnote{\ With\ presumed\ co-linearity\ of\ the\ position\ and\ acceleration\ vectors,\ never\ taking\ place,\ except again -- at apsides.} implies the annulling of the suitably  transformed transverse acceleration [$r\ddot{\varphi }+2\dot{r}\dot{\varphi }$ into $\dot{L}/r$], meaning 0-valued transverse acceleration, equating thus the radial and tangential ones -- a situation possible only for the rectilinear motion, apart from the circular motion with conventional center-directed `acceleration'.
	
	The reconsideration of the tenability of the Angular Momentum related Prime Integral should (have) be(en) important in particular that based on it has been developed the whole mechanics/dynamics of natural orbital systems, and subsequently analytical and rational mechanics, as well as the quantum mechanics, GTR and most recently the String Theory, with `spilling-over' into mathematics through the Symplectic Topology (``geometry of  `conservative' motion'') and the Symplectic Integration (actually, artificially ensuring the angular momentum constancy and thus the availability of the so-called ``Cyclic coordinate''). As it will be seen in the subsequent analysis and elaboration, the key role here has had the Kepler's transcendental equation relating angles to time and the resulting angular momentum (that is, the sectorial speed) constancy, and thereby the annulling of the suitably transformed transverse acceleration.
	
	Delving deeper into quite tedious long-term and indeed revolutionary Kepler's developments, and subsequent instigation of the unfortunately still ``uncompleted Copernican revolution'', reveals many relevant insights: from the Area law having initially been related to the mere substitute for calculating sums of products of distances ($r$) and the tangential (i.e. arc) velocities ($\upsilon $), traditionally being observed and fully relied on as invariant in the context of Ptolemaic Deferents and Epicycles with respect to Equants, which for Kepler was valid only for aphelion and perihelion, all way to the `extorted' transitioning from eccentric circle to elliptic trajectories -- the latter having been the most demanding, troublesome and time/efforts-consuming process, the final step of which was formulation of the famous Kepler's equation. 
	
	In this paper is shown that neither the Kepler's Second (Area) law nor its interpretation as the time-invariant angular momentum by far deserve the traditionally `fundamental' role in the modeling of natural orbital systems.  On the other side, Kepler's systematic adherence in contemplating the geometry underlying physical processes and the actual causes which imply a two-component central acceleration, i.e. the attractive-repulsive force (primarily based on Gibbs' insights exposed in his book \textit{De Magnete}), should have been given due attention and could very well have been leading towards development of consistent orbital mechanics and dynamics (in line with those of Descartes', Leibniz's, Boscovich's, and others'). 
	
	In the following, firstly in Section 2 is given outline of the historical development course by which Kepler had arrived at formulation of his first two laws. Then, in Section 3 is proceeded with the analysis of those results by analytic-numerical evaluation of the angular momentum and the transverse acceleration formulations, along the universally applicable truly oscillatory non-linear differential equations of motion supporting the Kepler's physical considerations, with further implications on mechanics, physics and mathematics being remarked on also in the concluding Section 4. 

\section{Historic Perspective of the Kepler's First Two Laws Development}
	
	Looked at in retrospect, the development of mechanics and physics reveals two `tracks' which branch off from Kepler: the official, over Newton, Maxwell, Einstein and further development of Quantum Mechanics, String Theory, etc., and the development that was to go over Descartes, Leibniz, Boscovich, Hertz and -- recently -- the mechanics and Aetherodynamics of Atsukovsky's rarefied-gas fluid, and Prigogine with the irreversible thermodynamics; thus, a direction which could (or at least faster and more plausibly) have lead to a scale-invariant unification of the phenomenological and 'substanti(ation)al' dynamics, and in them likely be found 'reflected' the ubiquitous Golden Section relationships, as well as the long sought for matching of the mathematical and physical continua.
	
	Kepler essentially did overcome both the Platonic precedence and supremacy of Geometry over the physical/empirical realm and the Aristotle's strict separation, that is  unrelatedness and/or incommensurateness between mathematics and physics, in that he actually did use them interchangeably: for one thing, the observations had to be representable by the `ideal' geometrical forms, and for the other -- those ideal geometrical forms had to be supportable by the viable physical mechanisms underlying the phenomena; Kepler thus became a founder of the 'physical mathematics', a discipline quite obviously in striking contrast to the contemporary ''mathematical physics'' {\dots} !?!
	
\begin{figure}[h!]
		\centering
		\includegraphics[width=11cm]{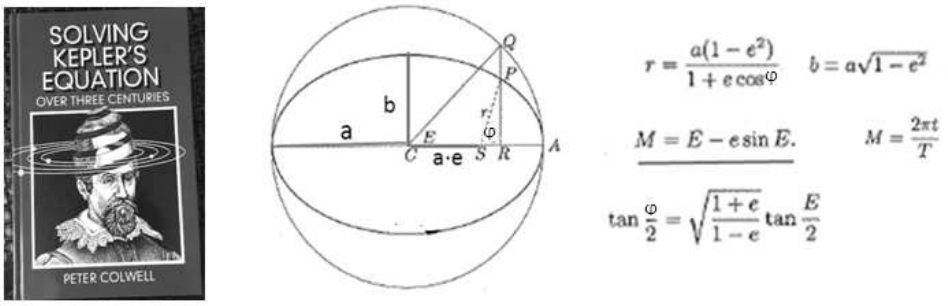}
		\caption{Set-up and final result of the Kepler's equation derivation; in continuation, in Fig. 2, use is made of positioning of the eccentric circle and/or ellipse rotated by 90 degrees, as in historic sources.}
		\label{fig:img1}
	\end{figure}

	The illustration on the cover page of the book on more than three centuries of solving the Kepler equation (by Peter Colwell, 1993, publisher: Willmann-Bell), shown in Figure \ref{fig:img1} (to the left), reflects extremely well the state and conditions (professionally and likewise, personally/privately) of Kepler himself during the first four very turbulent years\footnote{\ Instead\ of\ just\ one\ week,\ as\ he\ betted\ with\ the\ Tycho\ Brahe's\ estate\ manager\ --\ as\ per\ book\ of\ G.\ Badwill\ ``Kepler'',\ 1981.} of his work on determination of Mars' trajectory, and efforts of all those `giants' of science (from Newton to Cauchy and beyond) who were solving his transcendental Equation, not having been realizing that the `difficulties' were primarily related to not only relatively large values of the eccentricity factor, but rather to its insufficient physical adequacy, [5].
	
	Te Kepler's equation has traditionally been derived from the pre-established Area law, so that the pertinent relationship between the eccentric anomaly (E) and time is determined, as well as the relationship between the eccentric and the true ($\mathrm{\varphi}$) anomalies/angles, as reproduced in detail -- along the expressions displayed in this article, in corresponding appendices -- in the Appendix B \footnote{\ of\ the\ same\ titled\ article\ accessible\ at\ \url{ https://uns.academia.edu/SlobodanNedic/Papers} and/or transcripted preentations, or on the author's ResearchGate page \url{ https://www.researchgate.net/publication/337022429_Kepler's_Equation_and_Angular_Momentum_-_Historical_Perspective_Critical_Evaluation_and_Implications_for_Developments_in_Mechanics_and_Physics}}, whereas in Figure \ref{fig:img1} are given only the set-up and the end result, with the Kepler's equation being represented by the underlined part.     
	
	\noindent Immediately after the successful and very astute conversion of available astronomical measurements by Tycho Brahe from the Earth to the Sun as observation center, which meant overcoming 'retrogradeness' of the trajectories and aligning them with the Ecliptic [3], Kepler had gone a step back - to Eparches' Epicycles. In traversing all the steps - from eccentric circle, over the then analytically still intractable `oval', to the  ultimately adopted elliptic form - Kepler had finally, as after ``wakening from the Somnambulant dream'' - blissfully -- arrived at the sudden simultaneous validity of the distance law (otherwise, upfront considered as valid for only aphelion and perihelion) and the Area-law -- for the whole course of the elliptic trajectory\footnote{\ Subsequently,\ after\ certain\ criticisms\ of\ his\ result\ concerning\ the\ non-equidistancy\ of\ arcs,\ Kepler\ had\ explained\ that\ the\ Area-law\ (and\ thus\ the\ related\ constancy\ of\ the\ angular\ momentum)\ applies\ only\ to\ the\ `components'\ of\ arcs\ perpendicular\ to\ the\ radius,\ i.e. Sun's\  \textit{virtus motrix}, and not Mars'  \textit{vis insitia} --\ a\ `clarification'\ that\ unfortunately\ was\ not\ subsequently\ appreciated.}, although that was not so for the eccentric circle 'stage'.\\
	
\begin{figure}[h!]
		\centering
		\includegraphics[width=10cm]{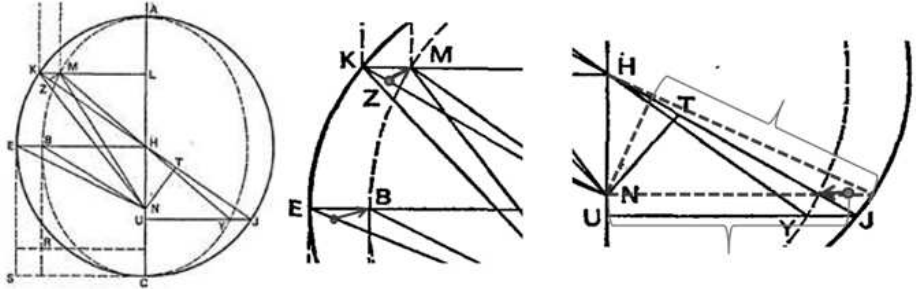}
		\caption{Modification conducted in the final stage of Kepler's derivations, [1].}
		\label{fig:img2}
	\end{figure}

	\noindent However, to accomplish this, he had undertaken - in retrospect, at least - an essentially inadmissible (but for him personally, in a number of ways - 'rescuing') step in suitably modifying the actual measurements, based on his reliance on ``diametral distance'' - projection of the distance from the focus to eccentric circle in direction of observation (K-N) on the line going through the related point at eccentric circle and its center, K-T - as indicated in Figure \ref{fig:img2}.
	
\section{Analytic-Numerical Critical Evaluation of Kepler's First Two Laws and Reaffirmation of His ''Physical Theory''}
By the formal identification of the swept area with the angular momentum $L$=$ r^{2} \dot{\varphi }$, i.e. their constancy for the coordinates' time-dependence uniquely, i.e. 'singularly' determined by Kepler's  Equation -- and only in that particular case, a priory - through the transformation $a_{p} =r\ddot{\varphi }+2\dot{r}\dot{\varphi }\Rightarrow \frac{1}{r} \frac{d}{dt} (r^{2} \dot{\varphi }) =\frac{1}{r} \dot{L} = 0$ -- gets annulled the transverse acceleration. 
	
	To check for this rather peculiar and by far -- due to the underlying modified measurements -- non-physical result, in the following is conducted the analytic-numeric evaluation of this, so that the time-dependence -- as of the distance \textit{r}, so also of the true anomaly $\varphi$ -- becomes explicated through time-dependence of the eccentric anomaly \textit{E}, the time-dependence of which is implied by Kepler's Equation.
	
\begin{figure}[h!]
		\centering
		\includegraphics[width=10cm]{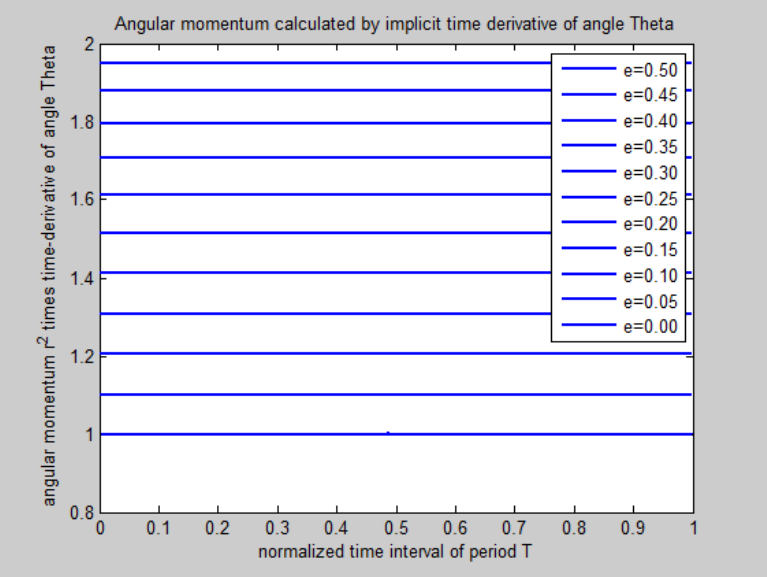}
		\caption{Analytic-numerical evaluation of the angular momentum expression.}
		\label{fig:img3}
	\end{figure}

	Firstly, in accordance with the set-up in Figure \ref{fig:img1}, along using the `alternative' parametric expression for ellipse $r(t)$=$a\cdot \left[1-e\cdot \cos \left(E(t)\right)\right]$, for the angular momentum is produced the following expression${}^{\ }$(with a=1+e)
     \begin{eqnarray}
	L = r^{2} \cdot \dot{\varphi }
	= (1+e)^{2} \frac{2\pi }{T} \sqrt{\frac{1+e}{1-e} } \cdot \frac{1-e\cdot \cos \left(E\right)}{\cos ^{2} \left(\frac{E}{2} \right)+\frac{1+e}{1-e} \cdot \sin ^{2} \left(\frac{E}{2} \right)} ,
    \end{eqnarray}
	and -- although not, or at least hardly recognizable by visual inspection -- it is constant\footnote{\ Actually,\ this\ is\ natural\ consequence\ of\ the\ positioning\ of\ the\ points\ on\ ellipse\ immediately\ bellow\ those\ on\ the\ eccentric\ circle\ and\ the\ subsequent\ determining\ the\ relationship\ between\ the\ angles\ and\ the\ time\ by\ 'forcing'\ the\ sameness\ of\ the\ subsequently\ swept\ elliptic\ sectors\ to\ equate\ those\ of\ the\ circle;\ it\ should\ be\ worth\ noting\ that\ this\ indicates\ importance\ of\ parametric\ elliptic\ and\ integration in\ general (to sum-up same 'amounts'),\ since\ only\ with\ such\ parameterization\ of\ angle\ the\ ellipse\ area\ becomes\ ab$\mathrm{\pi}$,\ and\ the\ corresponding\ table\ integral\ can\ be\ correctly\ evaluated,\ that\ is\ be\ possible\ to\ produce\ the\ sub-integral\ function\ by\ differentiation.} for all eccentricity factors \textit{e}, as shown in Figure \ref{fig:img3}, so that consequently the transformed transverse acceleration indeed is zero-valued. 

	\
	
\begin{figure}[h!]
		\centering
		\includegraphics[width=11.5cm]{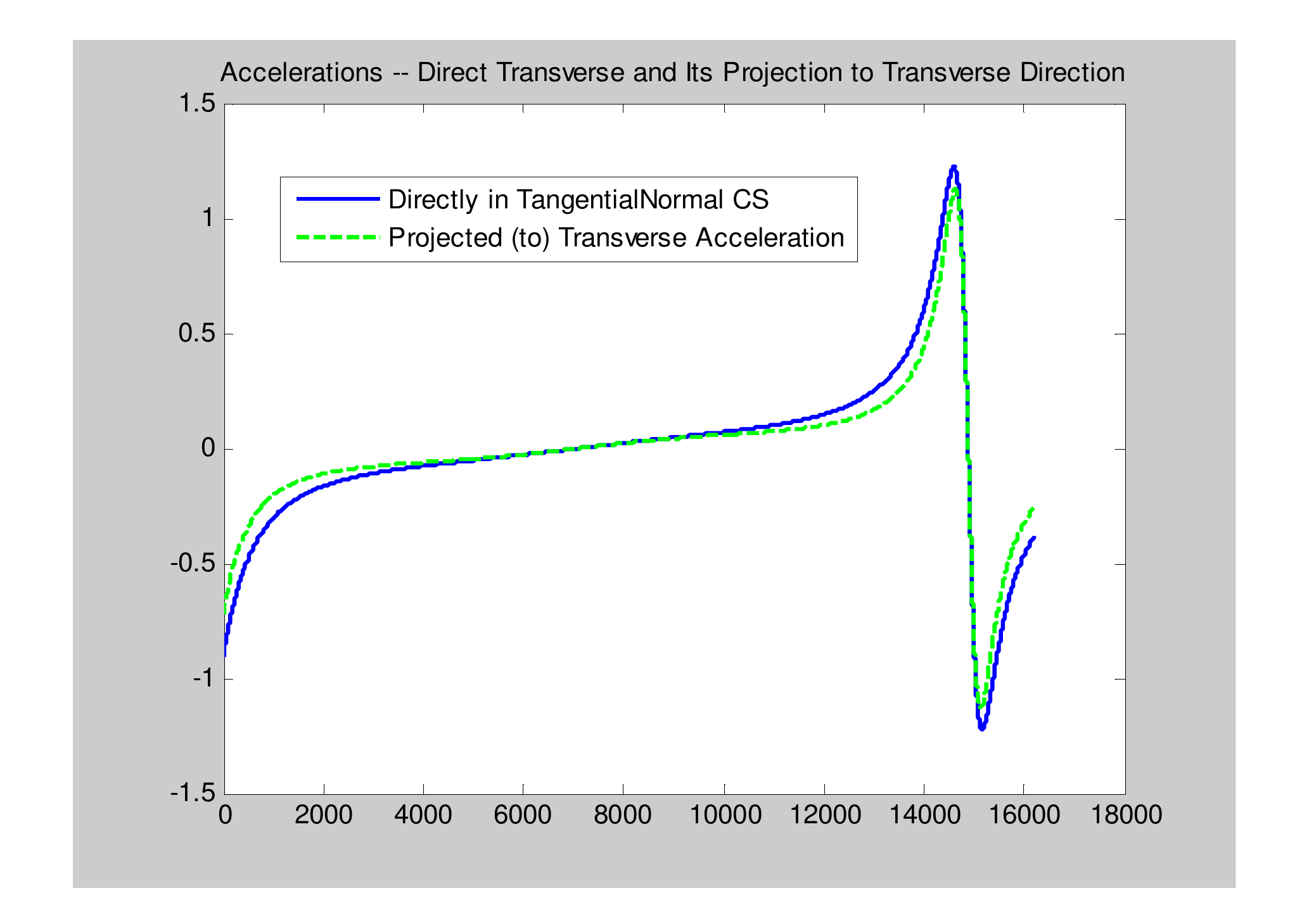}
		\caption{Analytic-numerical evaluation of the tangential and transverse accelerations.}
		\label{fig:img4}
	\end{figure}

	\noindent Indeed, the individually evaluated summing terms of the transverse acceleration form above 
	\\
	\begin{equation}
\begin{array}{l} 
r\cdot \ddot{\varphi} = 2\cdot \left(\frac{2\cdot \pi }{T} \right)^{2} \cdot a\cdot \sqrt{\frac{1+e}{1-e}} \cdot \frac{1}{\cos ^{2} (\frac{E}{2} )+\frac{1+e}{1-e} \cdot \sin ^{2} (\frac{E}{2} )} \cdot \frac{1}{\left[1-e\cdot cos(E)\right]} \cdot \left\langle \left[-\frac{e}{1-e} \right]\frac{\cos (\frac{E}{2} )\cdot \sin (\frac{E}{2} )}{\cos ^{2} (\frac{E}{2} )+\frac{1+e}{1-e} \cdot \sin ^{2} (\frac{E}{2} )} -\frac{e\cdot \sin (E)}{\left[1-e\cdot\cos(E)\right]} \right\rangle \end{array}
\end{equation}
\begin{eqnarray}
	2\cdot \dot{r}\cdot \dot{\varphi } = 2\cdot \left(\frac{2\cdot \pi }{T} \right)^{2} \cdot a\cdot \sqrt{\frac{1+e}{1-e} } \cdot \frac{e\cdot \sin (E)}{\cos ^{2} (\frac{E}{2} )+\frac{1+e}{1-e} \cdot \sin ^{2} (\frac{E}{2} )} \cdot \frac{1}{\left[1-e\cdot cos(E)\right]^{2} } ,
    \end{eqnarray}
	-- due to their essentially identical modules and differing signs -- maintain zero-valued transverse acceleration\footnote{\ Correction of the erroneous analytical derivation of the expression for the transverse acceleration was the reason for this article updating, and the change consists in changed sign of the second term in the angle-bracket in (2); due to the peculiar proportionality of the trigonometric expression determining the angular momentum constancy, these two terms have the same module, so with the wrong sign the related part was reduced to zero...}. However, when evaluating in the tangential-normal generalized coordinates one gets the non-zero tangential acceleration, the projection of which on the transverse direction produces non-zero transverse acceleration, as shown in Figure \ref{fig:img4}. 
	
\begin{figure}[h!]
		\centering
		\includegraphics[width=12.5cm]{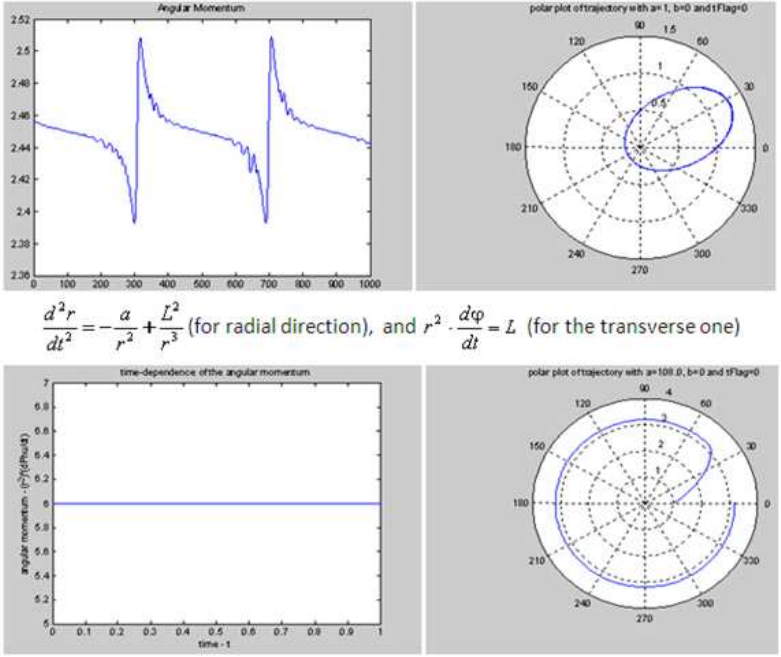}
		\caption{Numerical evaluation of the elliptic trajectory with unconstrained (upper part), and the angular momentum constrained to the very initial value (lower part; with (1/r)(d\textit{L}/dt)=0, instead of the second equation); the non-linear orbital differential equation is of the traditional, essentially non-oscillatory form, in which the radial equation has the so-called virtual or fictitious centrifugal force.}
		\label{fig:img5}
	\end{figure}

	It should be noted here that the numerical integration (in particular by the MATLAB function ode34.m), as shown in the two-part Figure \ref{fig:img5} of the conventional essentially non-oscillatory Newton-Laplaceian non-linear differential equations produces the elliptic trajectory with noticeably varying angular momentum, while by keeping the latter's constant (initially given) values at every integration step ultimately produces the circular orbit. By all means, this indicates non-physicality of the (angular momentum `taming') pertinent to Symplectic Integration.

\begin{figure}[h!]
		\centering
		\includegraphics[width=19cm]{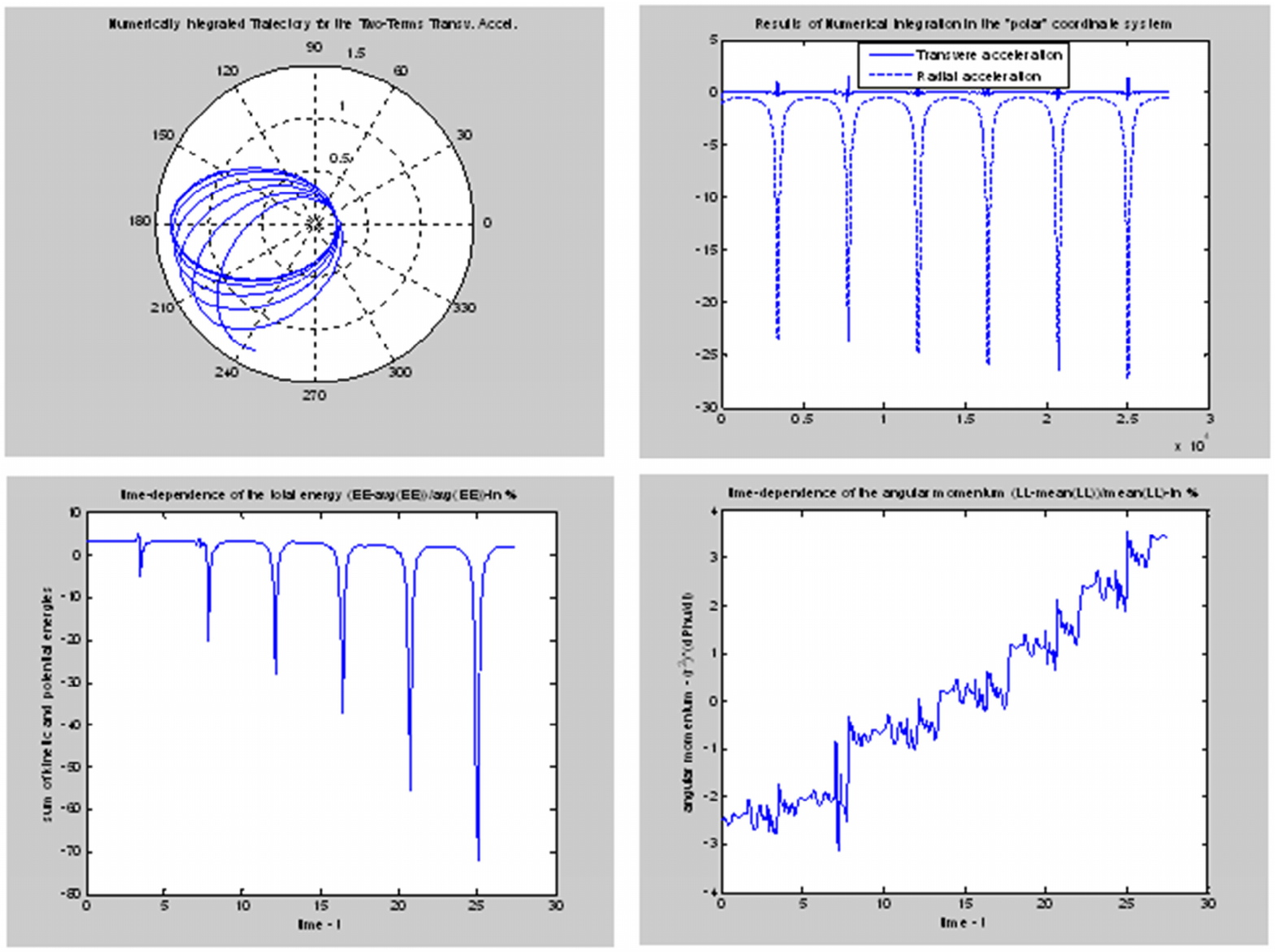}
		\caption{Results of numerical integration with two-component transverse acceleration and the angular momentum tamed to the initially given value in the radial equation; besides the trajectory, shown are the radial and transverse accelerations (upper, right), total energy variation (lower, left) and the effectively calculated - angular momentum (lower, right); note that the initial conditions are different from those in Fig. 5 are inconsequential for the effect exhibited.}
		\label{fig:img6}
	\end{figure}

    For the numerical integration with the complete two-terms expression for the transverse acceleration, which reproduces essentially the same elliptic trajectory as with the 'degenerated' azimuthal equation shown in Fig. 5, taming of the angular momentum in the radial equation impacts the excessive precessioning (as shown in Fig. \ref{fig:img6}, upper-left), casting a shadow of doubt on either validity of the Prime integral of angular momentum, or the sufficiency of just one component (attracting, without explicitly repulsing -- centrifugal -- part) in the central force. 

\begin{figure}[h!]
		\centering
		\includegraphics[width=16.5cm]{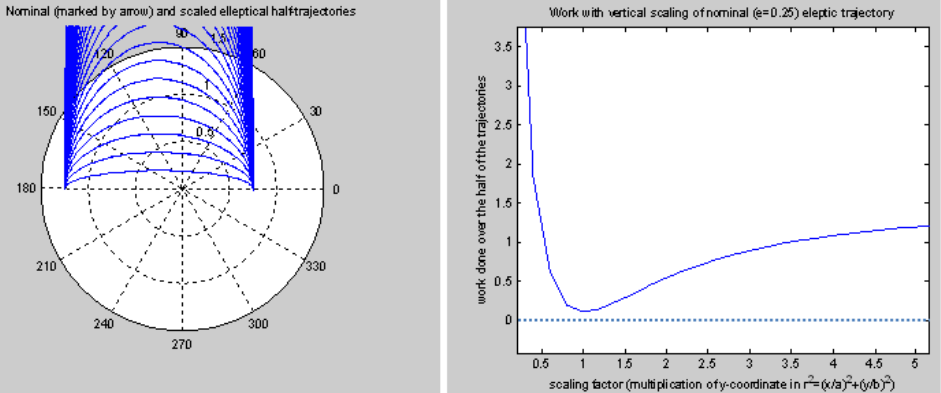}
		\caption{Numerical evaluation of the work done over the set of up- and down-scaled elliptic trajectories.}
		\label{fig:img7}
	\end{figure}

    As for the energy Prime integral implied zero-valuedness of the work done on a closed trajectory, the evaluation was undertaken of the work done over the set of trajectories produced by vertical scaling of the nominal elliptic trajectory with eccentricity factor e=0.25, to which the used Kepler's Equation applies (Figure \ref{fig:img7}), .

    \

    The result shown on the latter figure indicates the minimum of work done taking place on the nominal ellipse, the - kind of - positive energy well, suggesting the viable - and essentially largely plausible - determination of orbital trajectories by minimization of the work done\footnote{\ In\ producing\ these\ results,\ the\ absolute\ value\ of\ work\ is\ used,\ since\ the\ work\ is\ done\ during\ both\ the\ acceleration\ and\ the\ deceleration\ half-periods,\ which\ corresponds\ to\ "negative"\ and\ positive\ friction\ half-intervals,\ that\ is\ - the inflow and outflow of energy.}; thus, using the work done as action and directly minimising the related integral, instead of the traditional minimization of variation of actions determined by integration of, respectively,\ differences and sums of the kinetic and potential energy summation - in form of Lagrangeians and Hamiltonians. Although these implicitly themselves refute the validity of the First Energy Integral, the non-zero valued work done over an orbital elliptic trajectory directly overturns the tacit assumption of its zero-valuedness and the traditionally held conservativeness of the essentially time-varying (through the time-dependent distance) potential fields !?!   
	
	As justification for the Kepler's implicit (contemplated at the very first stage of transitioning to eccentric trajectory) two-component central acceleration, in line with the subsequent Leibniz's proposition $-\alpha /r^{2} +\beta /r^{3} $, can serve the -- introduced/proposed some century and a half ago -- Kepler-Ermakov system of non-linear differential equations [4], \noindent  $\frac{d^{2} r}{dt^{2} } -r\cdot \left(\frac{d\varphi }{dt} \right)^{2} $ = $-\frac{F(\varphi )}{r^{2} } +\frac{G(\varphi )}{r^{3} } $   and  $r\cdot \frac{d^{2} \varphi }{dt^{2} } +2\cdot \left(\frac{dr}{dt} \frac{d\varphi }{dt} \right)$ = $-\frac{dV(\varphi )/d\varphi }{r^{3} } $,                                        

    whereof exact invariant is $I $ = $ 0.5\left(r^{2} \dot{\varphi }\right)^{2} +V(\varphi )$, with $\eta $ = $1/r$, the orbital trajectory following from the equation
    \
	$h^{2} \left(\varphi ;I\right)\frac{d^{2} \eta }{d\varphi ^{2} } +h\left(\varphi ;I\right)\frac{\partial {\kern 1pt} h\left(\varphi ;I\right)}{\partial {\kern 1pt} \varphi } \frac{d\eta }{d\varphi } +\left(h^{2} \left(\varphi ;I\right)+F(\varphi )\right)\; \eta $ = $G(\varphi )$, along $h\left(\varphi ;I\right)$ = $\sqrt{2} \left(I-V(\varphi )\right)$. 

    The second term on the right-hand side of the first equation represents the explicit centrifugal force, and makes the conventional "fictitious" (suitably augmented second part of the geometric/kinematic summand of the radial acceleration by multiplying and dividing by the distance cubed, to contain the presumably constant angular momentum - as indicated in Fig. 7) centrifugal force - obsolete, at least. Here it would be important to note the presence of the explicit torque/ing term on the right hand side in the second (azimuthal-'direction') equation, directly effectuating the precessional motion. 

    Although in presence of the two-component explicit (attraction and repulsion) central forces in the first (radial-direction) equation this takes place even with this term be taken as equal to zero, it should/could have long been introduced as the projection of the radial central force on transverse direction also in the case of traditionally 'adopted' one-component (attractive) central force, instead of -- as exposed here -- its untenable, upfront annulling.
\
As final support to the genuine righteousness of Kepler's "Physical Astronomy" -- 'governed' by attraction and repulsion -- can serve the insight gained from the Atsukovsky's Aetherodynamics, in which inverse squared and inverse cubed distance dependencies turn out to be, respectively, electric and magnetic forces as manifestation of Bernoullianly interacting (respectively) equatorial and meridian flows pertinent to the toroidal vortex formations in the electromagnetic realm, and at all the scales of physical reality -- the subatomic and bellow and the galactic and above.     

\section{Concluding Remarks and Further Implications}
The main conclusion is that the First integral of Angular Momentum, and on it based notion of the related Cyclic Coordinate does not hold and should be abandoned altogether. Due mainly to his rather singular and entirely unphysical Equation, the subsequent developments of Orbital mechanics have made it as though Kepler - the mathematician had largely won over Kepler - the physicists. However, that obviously has been the misdeed of his `followers' and by far less his own\footnote{\ In\ the\ same\ way\ the\ inverse\ distance\ squared\ law\ of\ the\ gravitational\ attraction\ had\ been\ arrived\ at\ by\ combining\ the\ Huygen's\ centripetal\ acceleration,\ Kepler's\ Third\ law\ and\ the\ mass\ times\ acceleration\ force,\ Newton\ could\ as\ well\ have\ arrived\ at\ the\ inverse\ distance\ cubed\ centrifugal\ force\ by\ just\ using\ the\ Kepler's\ velocity\ times\ distance\ invariant\ at\ the\ aphelion\ and\ perihelion.}, in quite likely not thoroughly having read his works, thus not being aware of primarily the importance of the attracting-repulsing interactions, along his clarification of the true Area Law meaning, while mere development stages through which he was going, their `monumentality' as well as the demonstrated capability of overcoming his own pre-convictions and the initial presumptions are more than sufficient to align Kepler among the greatest of the `Giants''.    
	
	Along the untenability of the First Energy integral, the very deeply ingrained conservativeness and symmetry principles become largely obsolete and, for that matter, possibly also the whole areas of mathematics, as are the Symplectic Topology and Symplectic Integration. With support of the ubiquitous aetherodynamical interactions of the toroidal-vortex structures [2], one arrives at the unification of the electro-magnetic, gravitational-antigravitational, atomic -- weak and strong, inter-molecular, inter-stellar, inter-galac\-tical, etc. The Dark Matter then becomes just an untenable postulate brought up due to the deficiencies of the orbital mechanics/dynamics and in particular the untenability of the universal gravitational constant notion, while the Dark Energy becomes also such a postulate due primarily to unavailability of the explicit anti-gravitational (essentially - thermal, as the orbital body heat reduces the gravitational pressure gradient), repulsive mechanism. With the inter-relatedness of the Kepler-Ermakov non-linear differential equations, for modeling at the phenomenological level, and the Navier-Stokes equations, for the substantial level, along the Centre-Manifolds based modeling of the emergent complex dynamics systems [6] the basis is created for the scale-invariant and inter-scales interactions, and to arrive at the physical continuum which would be on pairs with the post-Cantorian set-theoretic mathematical continuum, [7]. As consequence, the alternative to the Big-Bang becomes the 'Grand-Recycling' in which out of substrate on lower scales -- through Entropy decrease -- get formed elements of the higher scale, and from the higher scales (for example out of the ``black-holes``) -- in the process of entropy increase --  the lower scales are getting replenished.
	
	With getting back some four centuries back and pursuing alternative sciences' development course, one thus can expect a profound change of the scientiffic and technological paradigm: 'pricipitativeness' along dissipativity, Morphysm along Entropysm, multiscaledness instead of muti-dimens\-ionality, unpredictable determinism instead of 'stohasticity', perpetuity of the 1st (with aether-energy) and 2nd instead of 3rd kind -- all towards the fullfiment of the 'ideal' of M. Petrovi\'{c} - Alas' Analogue Core and the  ``Dynamical Systems and Related Theories as the Algebra of Natural Laws\footnote{\ S.\ Ciulli's\ ``$  $Dynamical\ Systems\ and\ Microphysics:\ A\ Wish$  $,''$  $\ $  $2nd\ International\ Seminar\ on\ Mathematical\ Theory\ of\ Dynamical\ Systems\ and\ Microphysics,\ Udine,\ Italy,\ Sep\ 1-11,\ 1981.}\\
	
	\noindent\textbf{\textit{Acknowledgment:}}\textit{ Dedicated to the late Prof. Vujo Gordi\'{c} (originator of the Thermo - Dynamical Oscillator concept), and to the late Prof. Veljko Vuji\v{c}i\'{c} (originator of the Dynamics of Rheonomic Systems, arived at after proving the untenability of the First Integral of Energy).}\\
	
	\noindent \textbf{REFERENCES:}\\
	\\
	\noindent [1] Aiton E.J., The elliptical orbit and the area law, Section 10.5 in Vistas in Astronomy, Vol. 18, 1975 – Proceedings of works on occasion of Quartercentenary of Kepler’s birth
	
	\noindent [2] Atsukovsky A.V. , Obshchaya efirodinamika, IP RadioSoft, 2016; with translation to English of some of sections are available at 
	\href{https://www.dropbox.com/sh/b320a0is6fdc2e2/AADfjs\_LnbRBSaNXfxjPkTaqa?dl=00}{https://www.dropbox.com/sh/b320a0is6fdc2e2/AADfjs\_LnbRBSa\-NXfxjPkTaqa?dl=0}
	
	\noindent [3] Evans J., On the function and the probable origin of Ptolemy’s equant, Am. J. Phys. 52 (12), Dec. 1984
	
	\noindent [4] Leach P.G.L.  and Andriopoulus K., The Ermakov Equation: A commentary, Appl. Anal. Discrete Math, 2 (2008).
	
	\noindent [5] NASA, Orbital Flight Handbook, 1963.

    \noindent [6] Roberts A.J., Model Emergent  Dynamics in Complex Systems, May 25, 2018.

    \noindent [7] Sohrab S.H.,  “Continuum versus Quntum Fields Viewed Through a Scale Invariant Model of Statistical Mechanics,”  https://www.researchgate.net/publication/228757340\_Continuum\_versus\_Quantum\_Fields\_Viewed
    \_Through\_a\_Scale\_Invariant\_Model\_of\_Statistical\_Mechanics

\end{document}